\documentstyle[prl,aps,twocolumn]{revtex}
\begin{document}

\title{
 {\normalsize \hfill  TITCMT-95-2 }\\
 {~} \\
 Spectral function of the 1D Hubbard model in the $U\rightarrow +\infty $ limit
}

\author{
  Karlo Penc$^{a}$\cite{*}\cite{empenc},
  Fr\'ed\'eric Mila$^{b}$\cite{emmila} and Hiroyuki Shiba$^{a}$\cite{emshiba}
}

\address{
     $(a)$ Tokyo Institute of Technology, Department of Physics\\
           Oh-okayama, Meguro-ku, Tokyo 152 (Japan) \\
     $(b)$ Laboratoire de Physique Quantique, Universit\'e Paul Sabatier\\
     31062 Toulouse (France)\\
     {~} \\
\parbox[t]{14truecm}{\small
We show that the one-particle spectral functions of the one-dimensional Hubbard
model diverge at the Fermi energy like $|\omega-\varepsilon_F|^{-3/8}$
in the $U\rightarrow
+\infty $
limit. The Luttinger liquid behaviour $|\omega-\varepsilon_F|^\alpha$,
where $\alpha
\rightarrow 1/8$ as $U\rightarrow +\infty $,
should be limited to $|\omega-\varepsilon_F| \sim t^2/U$ (for $U$ large but
finite),
which shrinks to a single point,
$\omega=\varepsilon_F$, in that
limit.
The consequences for the observation of the Luttinger liquid
behaviour in photoemission and inverse photoemission experiments are discussed.
}}
\date{January 18, 1995}

\maketitle


\narrowtext

  Due to large quantum fluctuations, the low-energy physics of
interacting electrons in 1D is not of the Fermi
liquid type, but can be described by the Luttinger liquid theory
\cite{haldane,tomlut,solyom}. According to that theory,
the momentum distribution function should have no step at $k_F$ but should
behave like
$n_k-n_{k_F} \propto {\rm sign}(k_F-k) |k-k_F|^\alpha$, where $\alpha$ is a
non-universal exponent
that depends on the interaction. This has been confirmed for the
Hubbard model
by Ogata and
Shiba\cite{shiba} in the  $U\rightarrow +\infty $ limit,
in which case $\alpha = 1/8$\cite{alpha}.
Similarly, the
local (momentum averaged) one--particle spectral functions $A(\omega)$
(inverse photoemission spectrum) and $B(\omega)$ (photoemission spectrum)
defined by
\begin{eqnarray}
  A(\omega) &=&
 \sum_{f,\sigma}
  \left| \langle f,N+1| a^\dagger_{0,\sigma} |0,N\rangle \right|^2
  \delta(\omega \!-\! E^{N+1}_f \!+\! E^N_0) \nonumber\\
  B(\omega) &=&
 \sum_{f,\sigma}
  \left| \langle f,N\!-\! 1| a^{\phantom{\dagger}}_{0,\sigma} |0,N\rangle
  \right|^2
    \delta(\omega \!-\! E^{N}_0 \!+\! E^{N-1}_f) \label{eq:defaomega}
\end{eqnarray}
are expected to behave like
$|\omega-\varepsilon_F|^\alpha$ close to the Fermi energy.
In recent photoemission experiments, the function $B(\omega)$ has been
measured over a large energy range with a resolution of several meV \cite{exp}.
Under these
conditions, is it
possible to detect the Luttinger liquid behaviour? The
Luttinger liquid theory itself does not predict how far the power law
behaviour
will hold, and to answer this crucial question, a determination of the spectral
function for microscopic lattice models is necessary.
This is a very difficult problem. Even for the Hubbard model, which is soluble
by Bethe ansatz\cite{liebwu} and whose low--energy properties are reasonably
well understood\cite{woyna,frahm}, an exact calculation of the spectral
 functions
has not been possible so far,
although much has been done in this direction, both analytically
using bosonization\cite{specfunc}, canonical transformation\cite{oles} and
Bethe
Ansatz\cite{BA},
or numerically with Monte Carlo calculations\cite{montecarlo}.

 In this Letter, we present a calculation of
these spectral functions for the Hubbard model in the
$U\rightarrow +\infty $ limit.
The Hubbard model is defined by
\begin{equation}
  H= -t \sum_{i,\sigma}
   \left(
    a^\dagger_{i+1,\sigma} a^{\phantom{\dagger}}_{i,\sigma}
   + h.c. \right)
   + U \sum_{i} n_{i,\uparrow} n_{i,\downarrow}
\end{equation}
where $t$ is the hopping integral and $U$ is the on site repulsion. We will
 denote by
$N$ the number of fermions and
by $L$ the number of sites (we choose $L$ even),
the site index $i$ runs from $i\!=\!0$ to $L\!-\!1$.
Furthermore, we take $N$ to be of the form $4n+2$ ($n$ integer) so
that the ground-state is non--degenerate.
In the $U\rightarrow +\infty $ limit, it has been shown, using the Bethe Ansatz
solution, that the eigenstates can be written
as a product of a spinless
fermion wave--function
and a squeezed spin wave--function \cite{shiba}:
\begin{equation}
  |N,f \rangle = |\psi^{N}_{L,Q}(\{I\})\rangle
         \otimes |\chi^{N_\downarrow}_N(Q,\tilde f_Q) \rangle
\end{equation}
The spin wave--function $|\chi\rangle$ is characterized
by the number of down spins $N_\downarrow$, the total momentum $Q$, and
the quantum number
$\tilde f_Q$ within the
subspace of momentum $Q$. The spinless fermion part
$|\psi\rangle$ is an eigen--function of
$N$ spinless fermions on $L$ sites with momenta $k_j L = 2 \pi I_j + Q$, where
the  $I_j$, $j=1\dots N$, are integer quantum numbers.
The two components of the wave--function are coupled through
the momentum $Q$ of the spin wave function,  which imposes a
twisted boundary condition on the spinless fermion wave--function (each fermion
hopping from site $L-1$ to site $0$ will acquire a phase $e^{iQ}$).

In the limit $U\rightarrow +\infty$, all the states with different spin
configurations are degenerate
and the energy is equal to $-2 t \sum_j \cos k_j$, i.e. it does not depend on
the quantum numbers $\tilde f_Q$.
In the ground-state, the spinless fermion wave-function $|\psi_{GS,L}^N\rangle$
 is
described by the quantum numbers $Q=\pi$ and
$\{I\}=\{-N/2,\dots,N/2-2,N/2-1 \}$, so that the distribution of the $k_j$'s is
symmetric around the origin, the spin part being the ground-state of the
Heisenberg model according to Ogata and Shiba's prescription\cite{shiba}.

For a less
than half--filled model,
the spectral function $A(\omega)$ has contributions from both the lower and
upper
Hubbard
bands\cite{oles} and $A(\omega)=A^{LHB}(\omega)+A^{UHB}(\omega)$.
If we use the Ogata-Shiba wave function into Eq.~(\ref{eq:defaomega}), we get
$A^{LHB}(\omega)$ and $B(\omega)$. To do this,
we just write operator $a^\dagger_{0,\sigma}$ entering Eq.~(\ref{eq:defaomega})
as $b^\dagger_0 \hat Z^\dagger_{0,\sigma}$,
where  $b^\dagger_j$ creates a spinless fermion at site $j$ and
$\hat Z^\dagger_{j,\sigma}$ inserts a spin $\sigma$ after skipping the
first $j$ spins.
$A^{UHB}(\omega)$ can actually be obtained in that framework
by a minor modification of the wave-function\cite{longpaper}.
The summation over the different spin configurations ($\tilde f_Q$)
can then be performed because
the energy depends only on the quantum numbers $\{I\}$ and $Q$, and we get
\begin{eqnarray}
  A^{LHB}(\omega)
  &=&
 \sum_{Q,\sigma}
  C_{\sigma,N}(Q)
  A_{Q}(\omega) \nonumber\\
  A^{UHB}(\omega)
  &=& \frac{1}{L \!-\! N \!+\! 1}
 \sum_{Q,\tilde Q,\sigma}
  D_{N,-\sigma}(Q) B_{Q-\tilde Q}(U \!-\! \omega)\nonumber\\
  B(\omega)
  &=&
 \sum_{Q,\sigma}
  D_{\sigma,N}(Q)
  B_{Q}(\omega)
\end{eqnarray}
$C_{\sigma,N}(Q)$ and $D_{\sigma,N}(Q)$ are given by
\begin{equation}
  C_{\sigma,N}(Q) = \sum_{\tilde f_Q}
    \left| \langle \chi_{N+1} (Q,\tilde f_Q) |
      \hat Z^\dagger_{0,\sigma}
      | \chi_N(\pi,0) \rangle
    \right|^2
\end{equation}
with $Q=2 \pi j/(N+1)$ ($j$ integer) and
\begin{equation}
  D_{\sigma,N}(Q) = \sum_{\tilde f_Q}
    \left| \langle  \chi_{N-1}(Q,\tilde f_Q) |
      \hat Z_{0,\sigma}
      | \chi_{N} (\pi,0) \rangle
    \right|^2
\end{equation}
with $Q=2 \pi j/(N-1)$, while
$A_{Q}(\omega)$ is given by
\begin{equation}
 \sum_{\{I\}}
  \left|
    \langle \psi^{N+1}_{L,Q}(\{I\}) |
    b^\dagger_{0}
    | \psi^{N}_{GS,L} \rangle
    \right|^2
  \delta(\omega - E^{N+1}_f + E^N_0)
\end{equation}
and a similar definition holds for $B_Q(\omega)$.

The problem has now been reduced to the calculation of quantities involving
only the charge or the spin. A similar approach has already been followed by
Sorella and Parola\cite{sorella2} in their calculation of the momentum
distribution function. For instance,
our $D_{\sigma,N}(Q)$ is equivalent to their $Z(Q)$. However, they could not
calculate the spectral function away from half-filling because they did not
know how to
evaluate the charge part in that case. In the following, we show
how to calculate these quantities for any band-filling.

Let us start with $C_{\sigma,N}(Q)$ and $D_{\sigma,N}(Q)$.
These quantities satisfy the sum rules
$  \sum_Q C_{\sigma,N}(Q) = 1$ and
$  \sum_Q D_{\sigma,N}(Q) = N_\sigma/N $
and they have a singularity at $Q_\sigma= \pi N_{\sigma}/N$.
We have calculated them for small
clusters (Fig.~\ref{fig:cq}), and we found that for $N_\uparrow=N_\downarrow$,
in which case $Q_\uparrow = Q_\downarrow = \pi/2 \equiv Q_0$,
they
behave like $ \sim |Q-Q_0|^{-\eta}$ with $\eta=0.49\pm0.01$ if $Q>Q_0$
and $\eta=0.14\pm0.01$ if $Q<Q_0$ for $C_Q$ and $\eta=0.49\pm0.01$ if $Q<Q_0$
and $\eta=0.34\pm0.02$ if $Q>Q_0$ for $D_Q$. Let us note that,
although they are very small, $C_{\sigma,N}(Q)$ and $D_{\sigma,N}(Q)$
do not vanish
identically for $Q<Q_0$ and $Q>Q_0$ respectively, contrary to what was claimed
 in
Ref.\cite{sorella2}.
The exponents of the main singularities are consistent with
the theoretical value
1/2. In Fig.~\ref{fig:cq}, the solid lines are fits to the numerical results
that
have been used in the following.

Let us now turn to the charge part.
For $Q=\pi$, the only excited states that contribute are those with one
particle-hole pair, and  $A_{\pi}(\omega)$ is just the spinless fermion density
of states $1/\pi\sqrt{4t^2 -\omega^2}$.
The
problem is not so simple for $Q \ne \pi$ because we have
to evaluate matrix elements between states
with different boundary conditions. Let us suppose that $k$ and $k'$ correspond
to boundary conditions with $Q$ and $Q'$ respectively.
Then, one can easily show that:
\begin{equation}
    \left[ a^\dagger_{k'}, a^{\phantom{\dagger}}_{k} \right]_+
    = \frac{1}{L} e^{-i(k'-k)/2}e^{i(Q'-Q)/2}
       \frac{\sin ([Q'-Q]/2) }{\sin ([k'-k]/2)}
\end{equation}
Clearly, we are faced with Anderson's orthogonality
catastrophy\cite{ortho1} and states with
many particle---hole excitations will contribute.
More generally, the matrix element
$ \langle 0|
      a^{\phantom{\dagger}}_{k_N}
      \dots
      a^{\phantom{\dagger}}_{k_2}
      a^{\phantom{\dagger}}_{k_1}
      a^\dagger_{k'_1} a^\dagger_{k'_2}
      \dots
      a^{\phantom{\dagger}}_{k'_N}
   |0\rangle $
is given by
\begin{eqnarray}
 && L^{-N}
  e^{i(Q'-Q)N/2}
  \prod_j e^{-i (k'_j-k_j)/2}
  \sin^N\frac{Q'-Q}{2} \nonumber\\
&\times&  \left| \begin{array}{cccc}
      \sin^{-1} \frac{k'_1-k_1}{2} &
      \sin^{-1} \frac{k'_1-k_2}{2} &
      \dots &
      \sin^{-1} \frac{k'_1-k_N}{2}
    \cr
       & & &
    \cr
      \sin^{-1} \frac{k'_2-k_1}{2} &
      \sin^{-1} \frac{k'_2-k_2}{2} &
      \dots &
      \sin^{-1} \frac{k'_2-k_N}{2}
    \cr
       & & &
    \cr
      \vdots     &  \vdots     &       & \vdots
    \cr
       & & &
    \cr
      \sin^{-1} \frac{k'_N-k_1}{2} &
      \sin^{-1} \frac{k'_N-k_2}{2} &
      \dots &
      \sin^{-1} \frac{k'_N-k_N}{2}
   \end{array}
   \right|
  \nonumber
\end{eqnarray}
The central observation is that this
determinant is very similar to the Cauchy determinants\cite{ortho}
and that it can be expressed as a product:
\[
\pm
   \prod_{j>i} \sin \frac{k_j-k_i}{2}
   \prod_{j>i} \sin \frac{k'_j-k'_i}{2}
   \prod_{i,j} \sin^{-1} \frac{k'_i-k_j}{2}
\]
 where the sign is $+$ for $N=1,4,5,8,9,..$ and $-$ for $N=2,3,6,7,..$.
 After a straightforward calculation one finally
gets:
\begin{eqnarray}
  &&\left|
   \langle
     \psi^{N+1}_{L,Q}(\{I\}) | b^\dagger_{0}| \psi^N_{GS,L}
   \rangle
  \right|^2
  =
  \frac{1}{L^{2N+1}}
  \cos^{2N}\frac{Q}{2}
  \nonumber \\
  &\times&   \prod_{j>i} \sin^2 \frac{k_j-k_i}{2}
   \prod_{j>i} \sin^2 \frac{k'_j-k'_i}{2}
   \prod_{i,j} \sin^{-2} \frac{k'_i-k_j}{2}
  \label{eq:sinpro}
\end{eqnarray}
For the matrix element entering $B_Q(\omega)$, a similar expression holds
with $N$ replaced by
$N-1$.
With such an expression, it becomes possible to calculate the spectral
functions
numerically.
To do that, we generate
the quantum numbers $I_j$ and calculate the energy and the product of
Eq.~(\ref{eq:sinpro}).
It turns out that the sum rules
$
  \int_{-\infty}^\infty A_{Q}(\omega) d\omega = 1-N/L
$
and
$
   \int_{-\infty}^\infty B_{Q}(\omega) d\omega = N/L
$
hold separately for each $Q$.
 Moving from $Q=\pi$ to $Q=0$, the effect of the many--particle excitations
becomes important. The weight of the Van--Hove singularity decreases;
a power--law singularity appears near the
Fermi energy
and a tail appears beyond the Van--Hove singularity.
This tail comes from the incoherent part of $A_Q(k\sim\pi,\omega)$ and
$B_Q(k\sim 0,\omega)$, and its weight increases when $Q$ decreases.
Including up to three
particle-hole excitations, we already get almost all the spectral weight. For
instance, for $L=60$
we get 99.993\% of the total weight in the worst case, namely for $Q=0$. So,
for
all practical purposes, we can limit ourselves to three particle-hole
excitations.
This is an important observation because the number of
states grows exponentially with the size, and to take into account all the
states is
possible only for smaller systems ($L\approx30$).

We are now in a position to calculate numerically $A(\omega)$ and $B(\omega)$
and a
typical result is presented in Fig.~\ref{fig:specfunc}.
Note that the following sum-rules are satisfied \cite{sawatzky,oles}:
$
  \int_{-\infty}^\infty A^{LHB}(\omega) d\omega = 2(1-N/L)
$
and
$
  \int_{-\infty}^\infty A^{UHB}(\omega) d\omega =
  \int_{-\infty}^\infty B(\omega) d\omega = N/L
$ .

The most interesting and surprising result is that, instead of going to zero,
as
the Luttinger liquid theory predicts, the spectral functions increase when
$\omega \rightarrow \varepsilon_F$, where $\varepsilon_F=-2t\cos 2k_F$ is
the Fermi energy of the spinless fermions and $k_F=\pi N/2L$ is the Fermi
momentum of
the fermions with spin.
In fact, using the same framework, we can prove
analytically that they diverge in that limit.
To do that, we first use the fact that the low energy spectrum consists of
towers
of excitations centered at momenta $(N+1)Q/L+4\, p\, k_F$,
$p=0,\pm 1, \pm2,...$
to write $A_Q(\omega)$ as the sum of the contributions coming from each tower
$\tilde A_Q^p(\omega)$. The lowest excitation in tower $p$ corresponds to a set
of densely packed quantum numbers $I_j$ shifted by $p$.
 From the definition of the
momenta $k_j$, this is equivalent to imposing a twist of wave-vector
$Q+2 p \pi$,
so that $\tilde A_Q^p(\omega) = \tilde A_{Q+2 p \pi}^0(\omega)$.
So all we have to do is calculate
$\tilde A_{\tilde Q}^0(\omega)$, where $\tilde Q$ can take values inside and
outside the first
Brillouin zone. Now, $\tilde A_{\tilde Q}^0(\omega)$
has peaks at energies
$\varepsilon_0+ju_c2\pi/L$,
 $j=1,2,...$, where
$\varepsilon_0$ is the energy of the lowest peak in the tower. This energy is
equal to
$
E^{N+1}_0(\tilde Q) - E_0^{N} = \varepsilon_F
  + \pi u_c (1+\alpha_{\tilde Q})/L $,
where
\begin{equation}
  \alpha_{\tilde Q}=
     \frac{1}{2} \left( \frac{{\tilde Q}}{\pi} \right)^2 -
     \frac{1}{2}
\end{equation}
So $\tilde A_{\tilde Q}^0(\omega)$
can be written as
$\sum_j \tilde A_{\tilde Q}^j
 \delta (\omega\!-\!\varepsilon_0\!-\!ju_c2\pi/L)$.
 From Eq.~(\ref{eq:sinpro}), one can show that
\begin{eqnarray}
   \tilde A_{\tilde Q}^j & = &
     \frac{(1+\alpha_{\tilde Q})(2+\alpha_{\tilde Q})\dots(j+\alpha_{\tilde
Q})}
     {j!} \tilde A_{\tilde Q}^0
      +O(1/L)
    \nonumber\\
    &\approx& \frac{(j+1/2 +\alpha_{\tilde Q}/2)^{\alpha_{\tilde Q}}}
    {\Gamma(\alpha_{\tilde Q}+1)}
           \tilde A_{\tilde Q}^0       \label{eq:aj}
\end{eqnarray}
and
\begin{equation}
  \tilde A_{\tilde Q}^0 = \frac{1}{L} (L \sin \pi n)^{-\alpha_{\tilde Q}}
     f({\tilde Q}) +O(1/L)
\end{equation}
where the dependence on size and filling is taken care of.
 If we put all together, we get
\begin{equation}
  \tilde A_{\tilde Q}^0(\omega) \approx \frac{1}{2 \pi u_c}
    \frac{1}{\Gamma(\alpha_{\tilde Q}+1)}
    \left(\frac{\omega-\varepsilon_F}{2 \pi u_c \sin \pi n}\right)^
    {\alpha_{\tilde Q}}
    f({\tilde Q})
\end{equation}
The function $f({\tilde Q})$ for ${\tilde Q}\ll L$ satisfies the recursion
relation
\begin{equation}
   f({\tilde Q}+\pi) = f({\tilde Q}-\pi) \frac{\Gamma({\tilde Q}/2 \pi)^2}
   {\Gamma(-{\tilde Q}/2 \pi)^2}
                       \pi^{2{\tilde Q}/\pi}
\end{equation}
where $f(\pi)=1$ and in the
interval from $=0$ to $\pi$ it can be approximated within 0.01\% by
   \[
    \ln f({\tilde Q}) \approx -0.3047 + 0.3248\, {\tilde Q}^2/\pi^2 -
    0.0201 \, {\tilde Q}^4/\pi^4
   \]

Finally, due to the divergence of $C_{\sigma,N}(Q)$ and
$D_{\sigma,N}(Q)$ at $Q=\pi/2$,
$A^{LHB}(\omega)$
is dominated close to $\varepsilon_F$ by
$A_{\pi/2}(\omega)$. This function is itself dominated by
$\tilde A_{\pi/2}^0(\omega)$.
 which
diverges with the exponent $\alpha_{\pi/2} = -3/8$ at $\varepsilon_F$.
So, $A^{LHB}(\omega) \propto |\omega-\varepsilon_F|^{-3/8}$ near
$\varepsilon_F$. The whole proof can be reproduced for $B(\omega)$ which also
diverges as $|\omega-\varepsilon_F|^{-3/8}$. Let us note that this power law
cannot be clearly identified numerically with large but finite systems because
there are other subdominant diverging contributions.

This divergence of $A^{LHB}(\omega)$ with exponent $-3/8$ is different from
the prediction of the Luttinger liquid theory according to which
$A^{LHB}(\omega)$ should
vanish as $|\omega-\varepsilon_F|^\alpha$, where $\alpha$, the exponent that
 also
enters the momentum distribution function, is known to be equal to $1/8$ in the
$U\rightarrow +\infty$ limit. This apparent contradiction can be lifted as
follows. First, we note that the exponent $-3/8$ can be explained very simply
on
the basis of the Green's function of the large $U$ Hubbard model \cite{frahm}
which is given by
\[
 G(x,t) \sim
  \frac{e^{i (k_F x-\varepsilon_F t)}} {(x-u_c t)^{1/2}(x-u_s t)^{1/2}}
  \frac{1} {(x^2-u_c^2 t^2)^{1/16}}
\]
If we set $u_s=0$, then  $A(\omega)$ behaves like
\[
\int dt G(x=0,t) e^{i \omega t}
            \sim \int dt \frac{e^{i (\omega-\varepsilon_F) t}}{t^{5/8}}
            \sim (\omega-\varepsilon_F) ^{-3/8}
\]
while, if $u_s$ is finite, it behaves like
\[
\int dt G(x=0,t) e^{i \omega t}
            \sim \int dt \frac{e^{i (\omega-\varepsilon_F) t}}{t^{9/8}}
            \sim (\omega-\varepsilon_F) ^{1/8}
\]
However, in both cases, $n_k$ behaves like
\[
  \int dx G(x,t=0) e^{i k x}
            \sim \int dx \frac{e^{i (k-k_F)x} }{x^{9/8}}
            \sim |k-k_F|^{1/8}
\]
So, as long as $u_s>0$, or equivalently $U<+\infty$, the behaviour very
close to $\varepsilon_F$
is given by $(\omega-\varepsilon_F) ^{1/8}$, but this behaviour is
limited to frequencies $(\omega-\varepsilon_F)<u_s \sim t^2/U$.
In the limit $U\rightarrow
+\infty$, this domain shrinks to a single point, $\omega=\varepsilon_F$,
and the spectral
function diverges as $(\omega-\varepsilon_F) ^{-3/8}$. If $U$ is large but
finite, there will be a peak in the spectral function at $\omega \sim t^2/U$
reminiscent of this divergence.
This seems to be consistent with the Monte Carlo results
of Preuss {\it et al.} \cite{montecarlo} which show an increase of the spectral
 functions close to the Fermi level.

The implications for the experimental observation of the Luttinger liquid
behaviour are rather dramatic. To be able to see a difference from a step
function in spite of the
finite resolution, $\alpha$
should be large enough, which means one should consider strongly correlated
systems. But the present calculation shows that the range of validity of the
asymptotic law  $(\omega-\varepsilon_F) ^{\alpha}$ becomes very small if the
correlations are too big. Whether realistic models of one-dimensional
conductors
with intermediate values of the repulsion terms can lead to measurable effects
remains to be seen.

\begin{figure}
  \caption{
   $(N\!+\!1)C_{\sigma,N}(Q)$ (open symbols) and
   $(N\!-\!1)D_{\sigma,N}(Q)$ (solid symbols)
   for cluster sizes $N=10$, $14$, $18$, $22$, $26$ \label{fig:cq}}
\end{figure}

\begin{figure}
  \caption{$A^{LHB}(\omega)$ and $B(\omega)$ for the quarter--filled
  $U \rightarrow +
  \infty$ Hubbard model with L=260 and N=130.
  \label{fig:specfunc}}
\end{figure}

\end{document}